\documentclass[superscriptaddress,twocolumn,aps,showpacs,prl,amsmath,amssymb,floatfix]{revtex4-1}

\usepackage{amsmath,amssymb,color}

\usepackage{bm}
\usepackage{epsfig}
\usepackage{psfrag}
\usepackage[latin1]{inputenc}
\usepackage[english]{babel}
\usepackage{amsfonts}
\usepackage{amsmath}
\usepackage{upgreek}
\usepackage{hyperref}
\usepackage[dvipsnames]{xcolor}

\DeclareSymbolFont{bbold}{U}{bbold}{m}{n}
\DeclareSymbolFontAlphabet{\mathbbold}{bbold}

\begin{document}

\title{
	Soft Mode in the Dynamics of Over-realizable On-line Learning for Soft Committee Machines 
	}

\author{
Frederieke Richert}
\thanks{These  authors contributed equally}
\affiliation{
	Institut f\"{u}r Theoretische Physik, Universit\"{a}t
  Leipzig,  Br\"{u}derstrasse 16, 04103 Leipzig, Germany
	}
\author{
Roman Worschech}
\thanks{These authors contributed equally}
\affiliation{Max Planck Institute for Mathematics in the Sciences, D-04103, Leipzig, Germany}
\affiliation{
	Institut f\"{u}r Theoretische Physik, Universit\"{a}t
  Leipzig,  Br\"{u}derstrasse 16, 04103 Leipzig, Germany
	}

\author{
Bernd Rosenow}
\affiliation{
	Institut f\"{u}r Theoretische Physik, Universit\"{a}t
  Leipzig,  Br\"{u}derstrasse 16, 04103 Leipzig, Germany
	}	
\date{\today}

\begin{abstract}

Over-parametrized deep neural networks trained by stochastic gradient descent are successful in performing many tasks of practical relevance. One aspect of over-parametrization is the possibility that the student network has a larger expressivity than the data generating process. In the context of a student-teacher scenario, this corresponds to the so-called over-realizable case, where the student network has a larger number of hidden units than the teacher. 
For on-line learning of a two-layer soft committee machine in the over-realizable case, we find that the approach to perfect learning occurs in a power-law fashion rather than exponentially as in the realizable case. All student nodes learn and replicate one of  the teacher nodes if teacher and student outputs are suitably rescaled.

\end{abstract}
 
\maketitle
The research field of deep learning has recently attracted considerable attention due to significant progress in performing tasks relevant to many different applications  \cite{lecun2015deep,Goodfellow2016,Kri+12,Sil+17,Carleo+2019}. Neural networks are learning machines inspired by the structure of the human brain \cite{hebb1949organization}, which have been studied with methods from statistical mechanics \cite{hertz1991introduction,Wat_92+,Saad98,Engel2001,Carleo+2019,Bahri+20}, starting with simpler versions such as the perceptron \cite{Gardner88,Tishby90,Seung+92} and also including two-layer networks \cite{Schwarze+93,Opper94,biehl1995learning,riegler1995line,saad1995exact, saad1995line,Biehl+98,goldt2019dynamics,Goldt+20,Mei+18,ChiBa18,straat2019line}.
Often, learning is studied in the framework of the student-teacher scenario, in which a student has to learn the connection vectors according to which a teacher classifies input patterns \cite{Engel2001}.

One of the surprising properties of multi-layer neural networks is their ability to generalize well even in the over-parametrized regime, when the number of model parameters exceeds the number of training examples \cite{Zhang+2017}. 
Recently an understanding has started to emerge that generalization beyond the training data set can be successful even for strongly over-parametrized networks due to implicit regularization in a gradient descent based learning process \cite{Jacot+18,Arora+19,Belkin+19}. Bounds on the generalization error have been found to depend on  the size of the training data set according to a power law \cite{Arora+2019,CaGu20,Chen+20}, in contrast to the exponential decrease of the generalization error in a student-teacher scenario for soft committee machines in the realizable scenario \cite{saad1995exact,saad1995line}, where the number of hidden nodes is the same for student and teacher.

Motivated by the fact that in the over-parametrized regime the student network may have a larger expressivity than the process which generates the training data,  we study the evolution of the generalization error in the over-realizable case of learning in a student-teacher setup, where the student network has a  larger number of hidden units than the teacher network. 
For on-line learning (i.e.~one-pass stochastic gradient descent) of a soft committee machine, where each training example is presented to the student only once, it was found that  additional student nodes (beyond the number of teacher nodes) do not learn at all, i.e.~the weight vectors of the additional nodes decay to zero \cite{saad1995exact,saad1995line}. Only in a fully trained two-layer network has  learning of all student nodes been observed \cite{goldt2019dynamics,Goldt+20,Mei+18,ChiBa18}. Here, we present a rescaling of the output of soft committee machines such that all student nodes learn in the asymptotic limit of a large number of training data. Then, the  approach to perfect learning  is strikingly different as compared to the realizable case with an exponentially fast convergence to zero generalization error: convergence is of power-law type in the over-realizable case due to the presence of soft modes, which we demonstrate both numerically and analytically. 
In addition, for the case of a noisy teacher we present numerical evidence that the generalization error is smaller in the over-realizable case than in the realizable one (similar to the case of the fully trained two-layer network studied in \cite{goldt2019dynamics}). 

In our setup, both the student and the teacher network receive inputs $\boldsymbol{\xi}^{\mu}\in\mathbb{R}^N$ at time steps $\mu=1,...,p$ in the input layer, where the components are independent normally distributed ${\xi}^{\mu}_i \in \mathcal{N}\left(0,1\right)$. These inputs are processed in the hidden layer using a nonlinear mapping. The student network has $K$ hidden nodes, with the $k$-th node being characterized by the student vector $\boldsymbol{J}^{\mu}_k\in\mathbb{R}^N$ at time step $\mu$. The teacher network is similarly structured, having $M$ hidden nodes, with a teacher vector $\boldsymbol{B}_m\in\mathbb{R}^N$ associated with node $m$. A linear combination of the outputs of hidden units, obtained using a nonlinear activation function (in our case $g(x)=\text{erf}(x/\sqrt{2})$ due to its 	 analytic properties) yields the student output
\begin{equation}
\sigma(\boldsymbol{J}^{\mu},\boldsymbol{\xi}^{\mu})=\frac{\sqrt{M}}{K}\sum_{k=1}^{K}g\big(\boldsymbol{J}^{\mu}_k\cdot\boldsymbol{\xi}^{\mu}\big).
\end{equation}
Importantly, the normalization factor $\sqrt{M}/K$ is chosen such that it matches the teacher output $\zeta(\boldsymbol{B},\boldsymbol{\xi}^{\mu})=\frac{1}{\sqrt{M}}\sum_{m=1}^{M}g\big(\boldsymbol{B}_m\cdot\boldsymbol{\xi}^{\mu}\big)$, while at the same time ensuring that the over-realizability is attended to by the additional factor of $M/K$ as compared to the teacher normalization.

Learning of the student is achieved by minimizing the loss function $\epsilon(\boldsymbol{J}^{\mu},\boldsymbol{\xi}^{\mu})=\frac{1}{2}[\sigma(\boldsymbol{J}^{\mu},\boldsymbol{\xi}^{\mu})-\zeta(\boldsymbol{B},\boldsymbol{\xi}^{\mu})]^2$ in each time step $\mu$, using stochastic gradient descent \cite{biehl1995learning, riegler1995line}. In this procedure the $i$-th student vector is updated from time $\mu$ to time $\mu+1$ via the gradient of the loss function with respect to this particular student. The update rule for the $i$-th student vector with learning rate $\eta$ is
\begin{equation}
\boldsymbol{J}^{\mu+1}_i-\boldsymbol{J}^{\mu}_i=\frac{\eta}{N}\delta^{\mu}_i\boldsymbol{\xi}^{\mu} \ \ ,
\end{equation}
\begin{equation}
\delta^{\mu}_i=\frac{1}{K}g'\big(\boldsymbol{J}^{\mu}_i\cdot\boldsymbol{\xi}^{\mu}\big)\Big[\sum_{m=1}^M g\big(\boldsymbol{B}_m\cdot\boldsymbol{\xi}^{\mu}\big)-\frac{M}{K}\sum_{k=1}^K g\big(\boldsymbol{J}^{\mu}_k\cdot\boldsymbol{\xi}^{\mu}\big)\Big].
\end{equation}
In order to analyze the learning behavior of the student, we need to define the generalization error $\epsilon_g$, which is an average over all possible input vectors $\boldsymbol{\xi}$.
\begin{equation}
\epsilon_g(\boldsymbol{J}^{\mu})=\Big\langle \frac{1}{2}[\sigma(\boldsymbol{J}^{\mu},\boldsymbol{\xi}^{\mu})-\zeta(\boldsymbol{B},\boldsymbol{\xi}^{\mu})]^2 \Big\rangle_{\{\boldsymbol{\xi}\}}
\end{equation}
\begin{figure}[t]
	\centering
	\includegraphics[width=8.6cm]{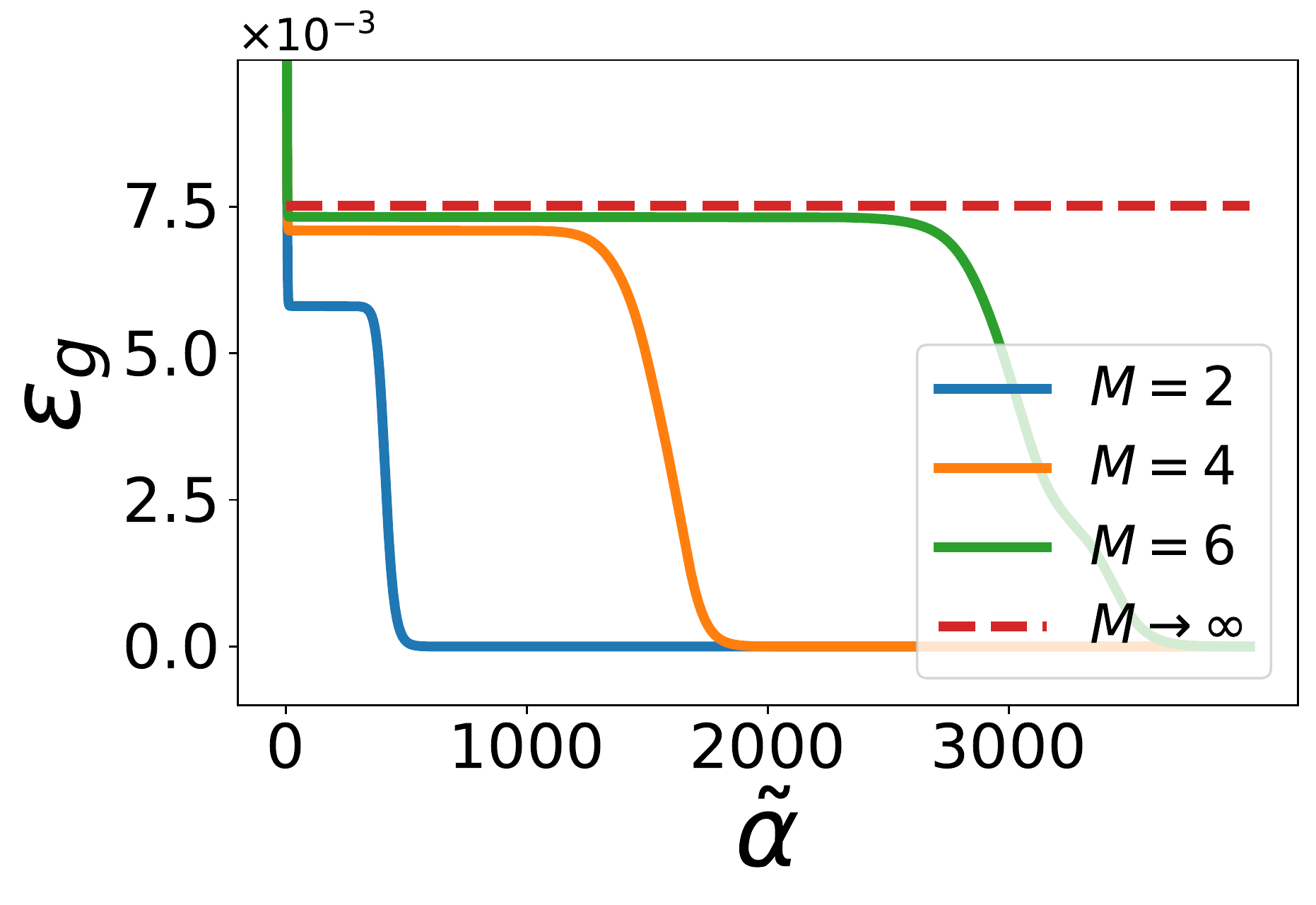}
	\caption{\label{Fig:Plateauheight} Dependence of the generalization error in the realizable scenario, $K=M$, on the normalized number of examples $\tilde{\alpha}=\frac{\alpha}{\eta}$ (``time''), computed to $O(\eta)$. Different colors indicate different widths $M$ of the teacher. The height of the plateau converges monotonously with $M$.}
\end{figure} 
\begin{figure}[h!]
	\centering
	\includegraphics[width=8.6cm]{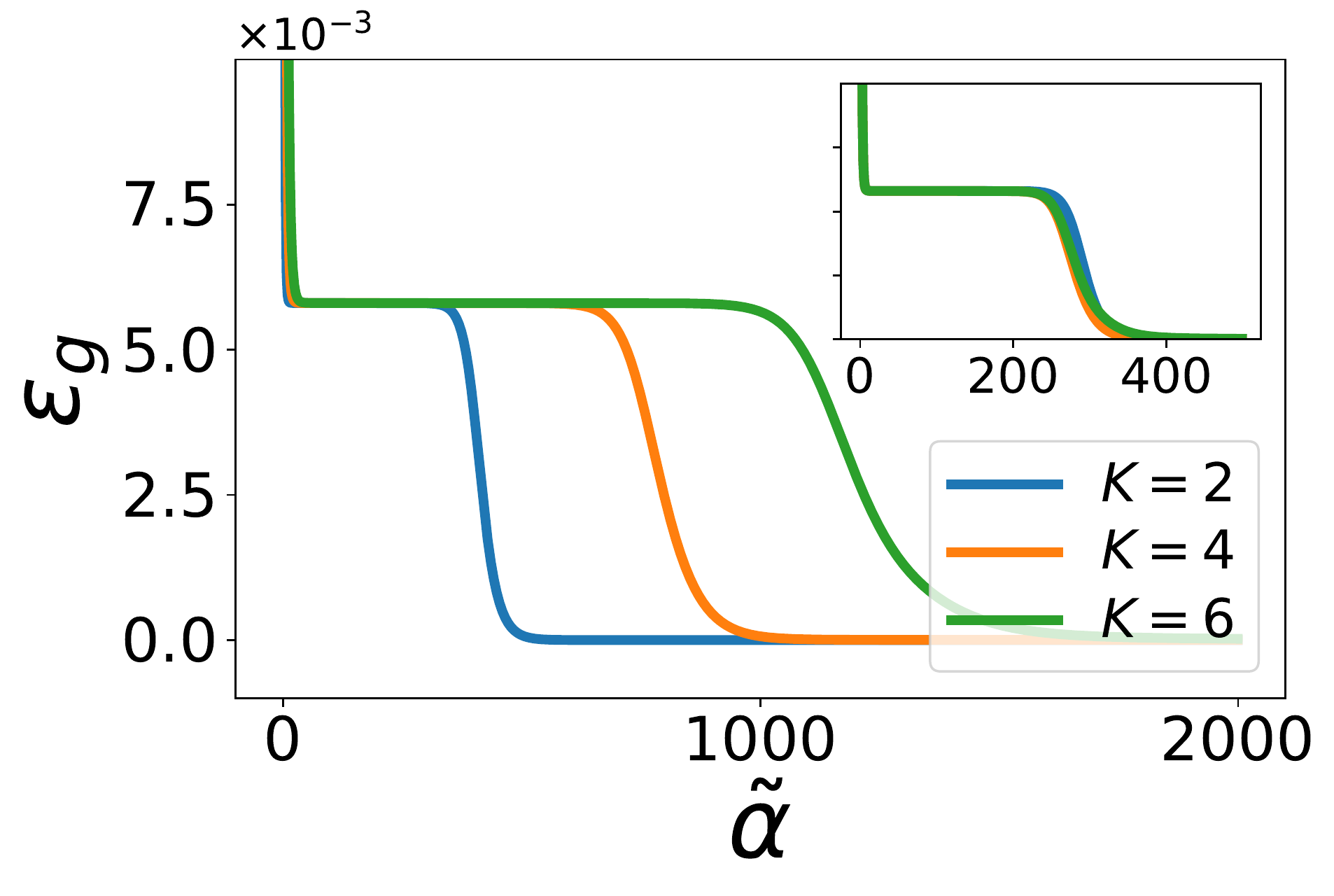}
	\caption{\label{Fig:Plateaulength} Evolution of the generalization error in realizable and over-realizable scenarios, $K\geq M$, 
	as a function of $\tilde{\alpha}=\frac{\alpha}{\eta}$, computed to $O(\eta)$ for $M=2$. The plateau height is independent of $K$, while the plateau length is proportional to $K$. The inset shows that a rescaling of the learning rate with $K$ leads to a collapse of the curves with different $K$.}
\end{figure} 
In order to compute this average in a statistical mechanics approach, we introduce order parameter matrices 
 $\mathbf{R}$, $\mathbf{Q}$, $\mathbf{T}$ with elements
$R^{\mu}_{in}= \boldsymbol{J}^{\mu}_i\cdot\boldsymbol{B}_n$, $Q^{\mu}_{ik}= \boldsymbol{J}^{\mu}_i\cdot\boldsymbol{J}^{\mu}_k$, $T_{mn}=\boldsymbol{B}_m\cdot\boldsymbol{B}_n=\delta_{mn}$,  with $i,k=1,...,K$ and $m,n=1,...,M$. They are the covariances of a multivariate Gaussian distribution, over which the averaging simplifies considerably \cite{saad1995line,Engel2001}. The generalization error thus becomes a function of these order parameters $\epsilon_g(\boldsymbol{R}^{\mu},\boldsymbol{Q}^{\mu})$. As we are interested in the typical behavior of the dynamics of our network, we take the thermodynamic limit $N\to\infty$, $p\to\infty$, where $\alpha=p/N$ stays finite \cite{saad1995exact, goldt2019dynamics}. Thus, we obtain a time evolution in the continuous time variable $\alpha$ for the student vector, and as a consequence, the dynamics of $\boldsymbol{R}(\alpha)$ and $\boldsymbol{Q}(\alpha)$ is given by \cite{saad1995exact}
\begin{subequations}
\begin{align}
\frac{d}{d\alpha} \boldsymbol{R}&=\eta\boldsymbol{F}(\boldsymbol{R},\boldsymbol{Q}) \label{dR/da} \\
\frac{d}{d\alpha}\boldsymbol{Q}&=\eta\boldsymbol{G}(\boldsymbol{R},\boldsymbol{Q})+\eta^2\boldsymbol{H}(\boldsymbol{R},\boldsymbol{Q}) \ \ .
\end{align}
\end{subequations}

The graph of the generalization error of soft committee machines has a characteristic shape \cite{saad1995exact}, as can be seen in Fig. \ref{Fig:Plateauheight}. The first structure of interest is the plateau of the generalization error, which corresponds to a plateau in the dynamics of $\boldsymbol{R}$ and $\boldsymbol{Q}$ \cite{biehl1995learning, saad1995line}. For this part of the time evolution the analytic ansatz $R_{in}\equiv R$, $Q_{ik}\equiv Q$ can be made, because all student vectors are found to have the same overlap with each other and with all the teacher vectors. In the small $\eta$ limit, i.e. neglecting $\eta^2$-terms, the equations for $R$ and $Q$ in this symmetric regime are
\begin{subequations}\label{plateauevolution}
\begin{align}
\frac{dR}{d\alpha}&=\frac{2\eta}{\pi}\frac{1}{1+Q}\frac{1}{K} \bigg\{\ \frac{1+Q-MR^2}{\sqrt{2(1+Q)-R^2}}-\frac{MR}{\sqrt{1+2Q}} \bigg\}\ \\
\frac{dQ}{d\alpha}&=\frac{4\eta}{\pi}\frac{1}{1+Q}\frac{1}{K} \bigg\{\ \frac{MR}{\sqrt{2(1+Q)-R^2}}-\frac{MQ}{\sqrt{1+2Q}} \bigg\}\ .
\end{align}
\end{subequations}
These equations show that with a proper normalization of student and teacher output, we achieve an effective scaling of the learning rate with $1/K$, which leads to interesting further results. The fixed points of \eqref{plateauevolution} are $R^*=(M(2M-1))^{-1/2}$, $Q^*=(2M-1)^{-1}$, the same as in the realizable case \cite{saad1995line}. But the length and height of the plateau in $\epsilon_g$ are interesting: The escape time of the symmetric plateau is proportional to the learning rate \cite{Biehltransient}, leading to a prolongation of the plateau by a factor of $K$ in the over-realizable case. The height of the generalization error in the plateau region is given by
\begin{equation}
    \epsilon_g^*=\frac{1}{6}-\frac{1}{\pi}M \arcsin\Big(\frac{1}{2M}\Big),
\end{equation}
which converges to $\lim_{M\to\infty}\epsilon_g^*=(1/6-1/2\pi) \approx 0.0075$ in the limit of an infinitely wide teacher network, independent of the number of student hidden nodes $K$. 
\begin{figure}[t]
	\centering
	\includegraphics[width=8.6cm]{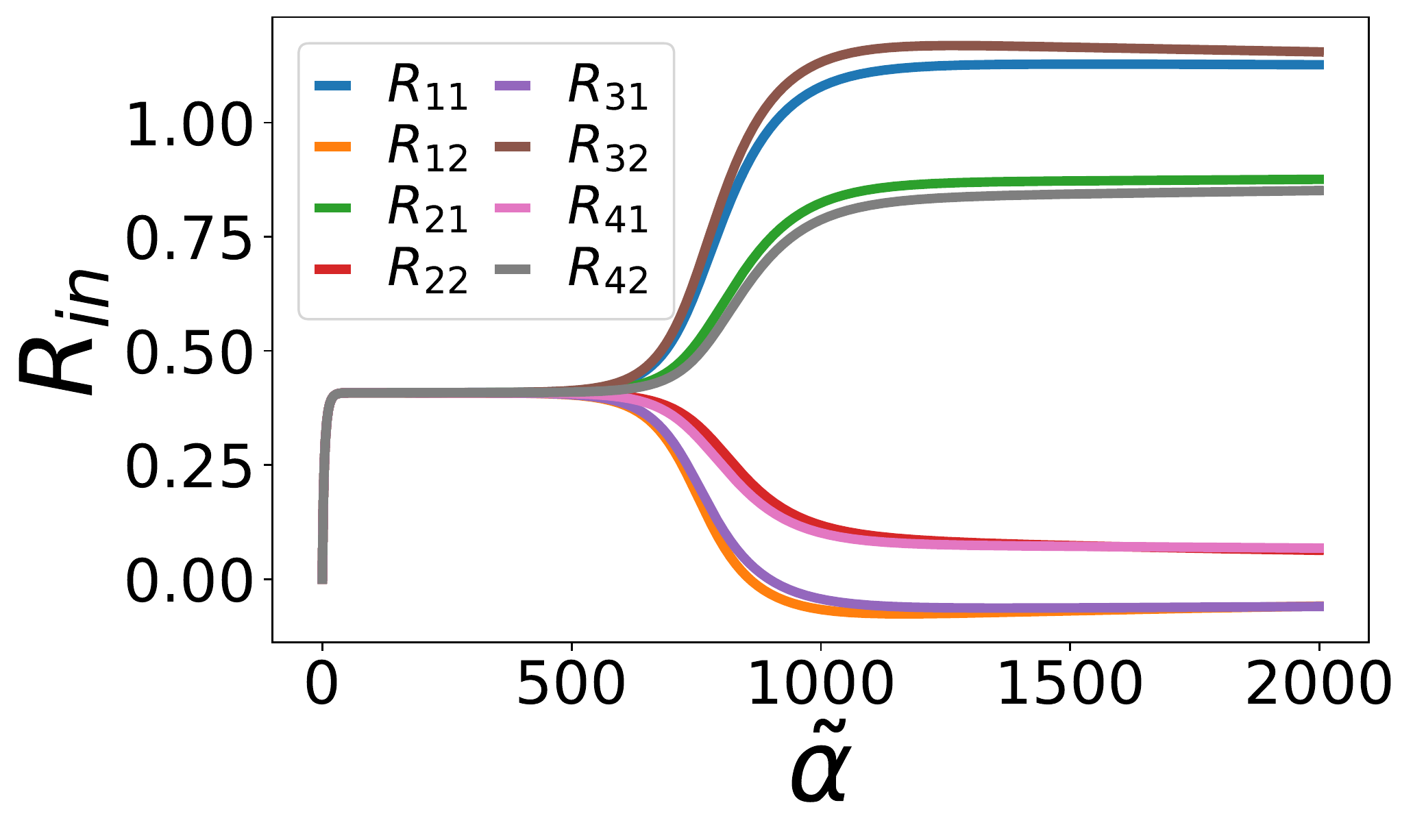}
	\caption{\label{Fig:R} Evolution of the student-teacher overlap $R_{in}$ as a function of $\tilde{\alpha}=\frac{\alpha}{\eta}$, computed to $O(\eta)$ for $M=2$, $K=4$.   Students 1 and 2 imitate teacher 1 after the specialization transition, while student 3 and 4 imitate teacher 2, indicated by the $R_{in}$ being close to one. The remaining overlaps tend to zero.}
\end{figure} 
These analytical results agree well with numerical simulations: Fig.~\ref{Fig:Plateauheight} shows the increasing plateau height for increasing dimensions $M$ of the teacher, which converges to the analytically derived value. In addition, the value of $\epsilon_g^*$ does not depend on the dimension of the student $K$, as can be seen in Fig.~\ref{Fig:Plateaulength}, where one can also observe that the plateau length does depend on $K$ in a linear fashion. We also investigated the correlation between plateau length and the initialization of $\boldsymbol{R}$. It turns out that the plateau length is inversely proportional to the logarithm of the variance taken for the initialization of the student-teacher overlaps. 

In the symmetric regime, all student vectors are found to behave similarly, in the sense that they have the same overlap with a given teacher vector. An exit from this region is achieved via the so called specialization transition \cite{saad1995exact,saad1995line, biehl1995learning}. As the name indicates, the student vectors now start to imitate one particular teacher vector each. We focus on the over-realizable scenario ($K>M$), such that there are more student vectors than teacher vectors. For $Z:= K/M\in \mathbb{N}$, our model, due to its particular normalization, allows all student vectors to learn, differing from the previous finding that $K-M$ student vectors are redundant and reduce their length to zero \cite{saad1995line}.

\begin{figure}[t]
	\centering
	\includegraphics[width=8.6cm]{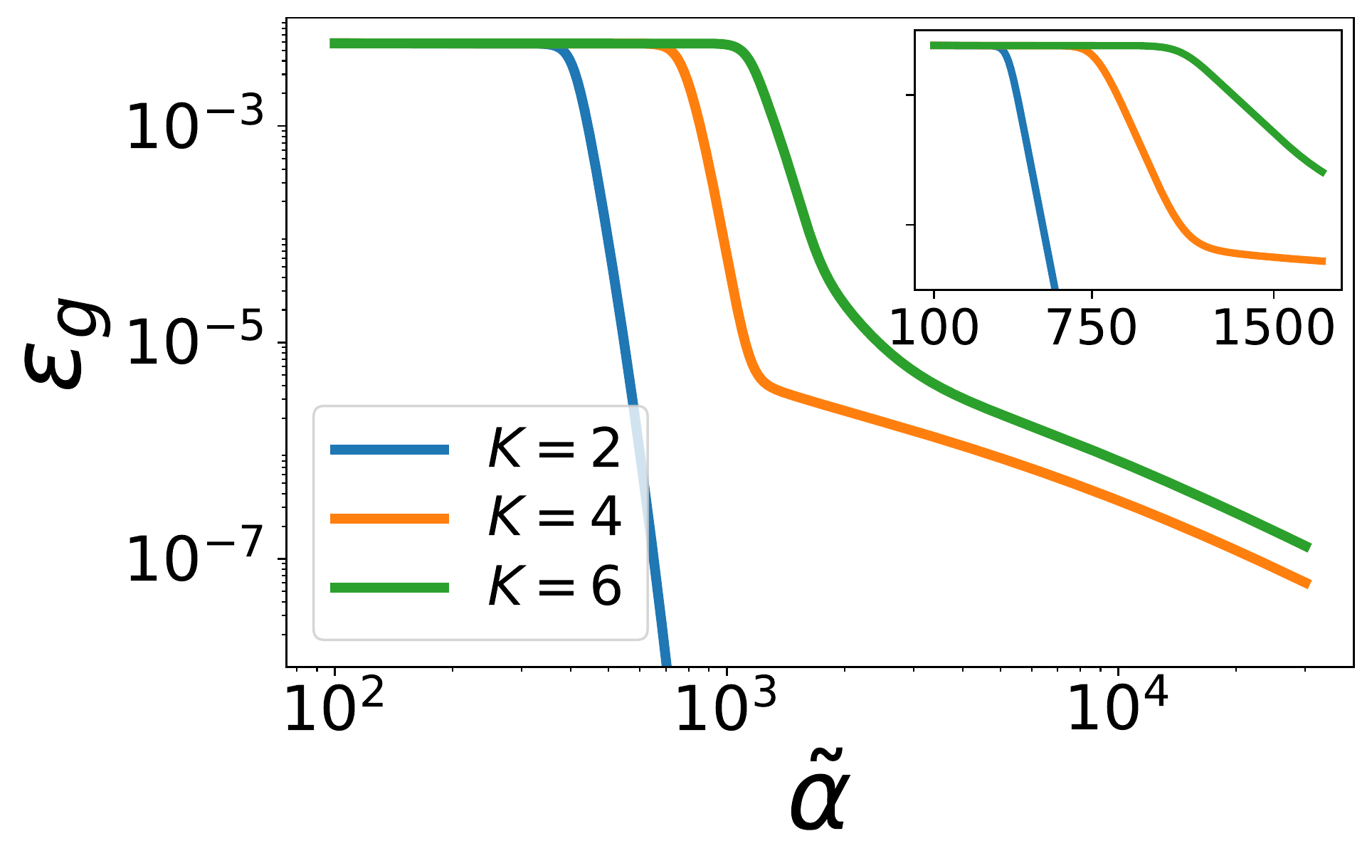}
	\caption{\label{Fig:eps1}   Double logarithmic plot of the asymptotic generalization error as a function of $\tilde{\alpha}=\frac{\alpha}{\eta}$, computed for $M=2$ to  $O(\eta)$,  in both the  realizable $K=M$ and the over-realizable scenario $K\geq M$. The inset shows the evolution of the generalization error on a single logarithmic scale during the early stage of the specialization transition. 
	While the generalization error approaches perfect learning with an exponential decay in the realizable case, one finds an exponentially decaying transient  followed by a power law $\epsilon_g \propto \tilde{\alpha}^{-2}$  for $K>2$.
	 }
\end{figure}

This behavior is reflected in the asymptotic regime of the generalization error, which we examine both numerically and analytically. Numerically, we find an algebraic convergence of the asymptotic evolution of both the order parameters $\boldsymbol{R}$, $\boldsymbol{Q}$ and the generalization error $\epsilon_g$ (see Figs.~\ref{Fig:R} and \ref{Fig:eps1}, respectively). Fig.~\ref{Fig:R} shows the evolution of the overlaps $R_{in}$ between student vector $i$ and teacher vector $n$. One can observe that there is a plateau regime, where all the overlaps are equal. After the specialization transition the overlaps of the first two student vectors with the first teacher vector and the overlaps of the third and fourth student vector with the second teacher vector are in the vicinity of one, while the remaining overlaps slowly approach zero. Interestingly, the overlaps of two student vectors imitating the same teacher vector add up to one, which will be exploited in our analytical solution. Fig.~\ref{Fig:eps1} shows the generalization error in the asymptotic regime for different scenarios. From the inset it can be seen that the exponential decay of the realizable case \cite{biehl1995learning, saad1995line} is replaced by a slower, power-law rate of convergence in the over-realizable case, with $\epsilon_g \propto1/\alpha^2$.\\
Inspired by the numerical results for the asymptotic evolution of our dynamical system, we investigate the case $M=2$, $K=4$ using the ansatz $ \boldsymbol{J}_1 = (1+\lambda_1)\mathbf{B}_1 +\lambda_2\mathbf{B}_2$,
$\boldsymbol{J}_2 = (1+\lambda_3)\mathbf{B}_1 +\lambda_4\mathbf{B}_2$, with $\{\lambda_j\}_{j=1}^4$ being small parameters. Thus, the first two student vectors both try to imitate the first teacher vector, while the third and fourth student vectors attempt to imitate the second teacher vector, described by an analogous ansatz. We thus obtain a four parameter ansatz for the asymptotic regime, with $\boldsymbol{R}$ and $\boldsymbol{Q}$ being of block matrix structure
\begin{align}
    \boldsymbol{R}=\left(
    \begin{array}{cc}
     1+\lambda _{1} & \lambda _{2} \\
     1+\lambda _{3} & \lambda _{4} \\
     \lambda _{2} & 1+ \lambda _{1} \\
     \lambda _{4} & 1+ \lambda _{3} \\
    \end{array}
    \right) 
    \hspace{1.5cm}
    \boldsymbol{Q}=\left(
\begin{array}{cc}
\boldsymbol{Q}_1 & \boldsymbol{Q}_2 \\
\boldsymbol{Q}_2 & \boldsymbol{Q}_1 
\end{array}
 \right). 
\end{align}
\indent Due to the linear dependence of $\boldsymbol{R}$ on the parameters $\lambda_j$, we can express \eqref{dR/da} as dynamical equations in terms of a vector $\boldsymbol{\lambda}=(\lambda_1,\lambda_2,\lambda_3,\lambda_4)^T$, yielding $\frac{d}{d\alpha}\boldsymbol{\lambda}=\eta\boldsymbol{F}(\boldsymbol{\lambda})$. In order to gain insight into the dynamics of the system, we linearize $\eta\boldsymbol{F}(\boldsymbol{\lambda})$, obtaining $\frac{d}{d\alpha}\boldsymbol{\lambda}=\boldsymbol{L}\boldsymbol{\lambda}$. We find that the matrix $\boldsymbol{L}$ has eigenvalues $\left\{0,0,l_+,l_-\right\}$, with $l_{\pm}=\frac{\left(\pm\sqrt{129}- 8 \sqrt{3}\right) \eta }{72 \pi} <0$. In contrast to the realizable case studied earlier \cite{biehl1995learning, saad1995line}, where exponential convergence was found, we obtain a doubly degenerate eigenvalue zero, hinting at a slower mode of convergence. The two eigenvectors corresponding to these modes are found to be $v_1=(0,-1,0,1)^T$ and $v_2=(-1,0,1,0)^T$.

In order to make further progress, we perform a coordinate transformation using the transpose of the eigenvector matrix. Exploiting the symmetry of $\boldsymbol{L}$, we introduce the transformation $\boldsymbol{\Lambda}= \boldsymbol{O}\boldsymbol{\lambda} $, which diagonalizes $\boldsymbol{L}$. The dynamical system in the new coordinates $\{\Lambda_j\}_{j=1}^4$ is then given by \begin{equation}
\frac{d}{d\alpha}\boldsymbol{\Lambda}= \boldsymbol{O} F(\boldsymbol{O}^{-1}\boldsymbol{\Lambda}).
\end{equation}\\
This is still a coupled system of ordinary differential equations, which we expanded to third order in the parameters $\Lambda_j$. The system of equations we solved is of the form
\begin{subequations}\label{Lambdaevo}
\begin{align}
\label{Lambda_1evo}\frac{d}{d\alpha}\left(
\begin{array}{c}
\Lambda_1\\
\Lambda_2\\
\end{array}
\right) &= \boldsymbol{F}_1(\Lambda_1^3,\Lambda_1^2\Lambda_2,\Lambda_1\Lambda_2^2,\Lambda_2^3)\\ 
& + \boldsymbol{F}_2(\Lambda_1\Lambda_3,\Lambda_1\Lambda_4,\Lambda_2\Lambda_3,\Lambda_2\Lambda_4)
\nonumber\\
\frac{d}{d\alpha}\left(
\begin{array}{c}
\Lambda_3\\
\Lambda_4\\
\end{array}
\right) &= 
\left(
\begin{array}{c}
l_+\Lambda_3\\
l_-\Lambda_4\\
\end{array}
\right)+
\boldsymbol{F}_3(\Lambda_1^2,\Lambda_1\Lambda_2,\Lambda_2^2)\label{Lambda_3evo}.
\end{align}
\end{subequations}
\indent As a first guess for solving these equations, we assume that the exponential convergence of the fast modes $\Lambda_3$, $\Lambda_4$ is dominant, such that we can set $\Lambda_3=0=\Lambda_4$. Then, \eqref{Lambda_1evo} transforms into an equation of the form $\frac{d}{d\alpha}f=\hat{c} f^3$ for $\Lambda_1$, $\Lambda_2$, solved by $f=\frac{\hat{c} }{\sqrt{\alpha}}$. Inserting this ansatz for $\Lambda_1$, $\Lambda_2$ into \eqref{Lambda_3evo}, we obtain the differential equation $\frac{d}{d\alpha}\Lambda_{3/4}=l_{\pm}\Lambda_{3/4}+\frac{\hat{c} _{3/4}}{\alpha}$, which is solved by $\Lambda_{3/4}= (\hat{c} _{3/4}  Ei(-l_{\pm}\alpha)+\Tilde{c}_{3/4} )e^{l_{\pm}\alpha }$, where $Ei$ is the standard exponential integral function. Studying the asymptotics of this function, one finds a convergence to zero as $\Lambda_{3/4}\propto\frac{1}{\alpha}$. To realize a self-consistent solution of \eqref{Lambdaevo}, we at last introduce also the coupling term $\boldsymbol{F}_2$, which has only the effect of changing the constants in the solutions for $\Lambda_1$, $\Lambda_2$. Thus, we have found an asymptotic solution $\Lambda_{1/2}=\frac{\hat{c}_{1/2}}{\sqrt{\alpha}}$, $\Lambda_{3/4}=\frac{\hat{c}_{3/4}}{\alpha}$.\\
To obtain the solution in our original parameters $\lambda_j$ we perform the coordinate back-transformation, yielding
\begin{equation}
    \boldsymbol{\lambda}=
    \left(\begin{array}{c}
    \frac{c_1}{\alpha}+\frac{c_2}{\sqrt{\alpha } }\\
    \frac{c_3}{\alpha}+\frac{c_4}{\sqrt{\alpha } }\\
    \frac{c_1}{\alpha }-\frac{c_2}{\sqrt{\alpha }}\\
    \frac{c_3}{\alpha}-\frac{c_4}{\sqrt{\alpha } }
    \end{array}\right).
\end{equation}
\indent This result enables us to compare the analytical findings for the asymptotic behavior of $\boldsymbol{R}$ directly to the numerical findings, with excellent agreement. Inserting our solution into the $\boldsymbol{R}$ and $\boldsymbol{Q}$ matrices and determining the generalization error, we find $\epsilon_g  \propto 1/\alpha^2$, in agreement with the numerical results discussed earlier.
\begin{figure}[t]
	\centering
	\includegraphics[width=8.6cm]{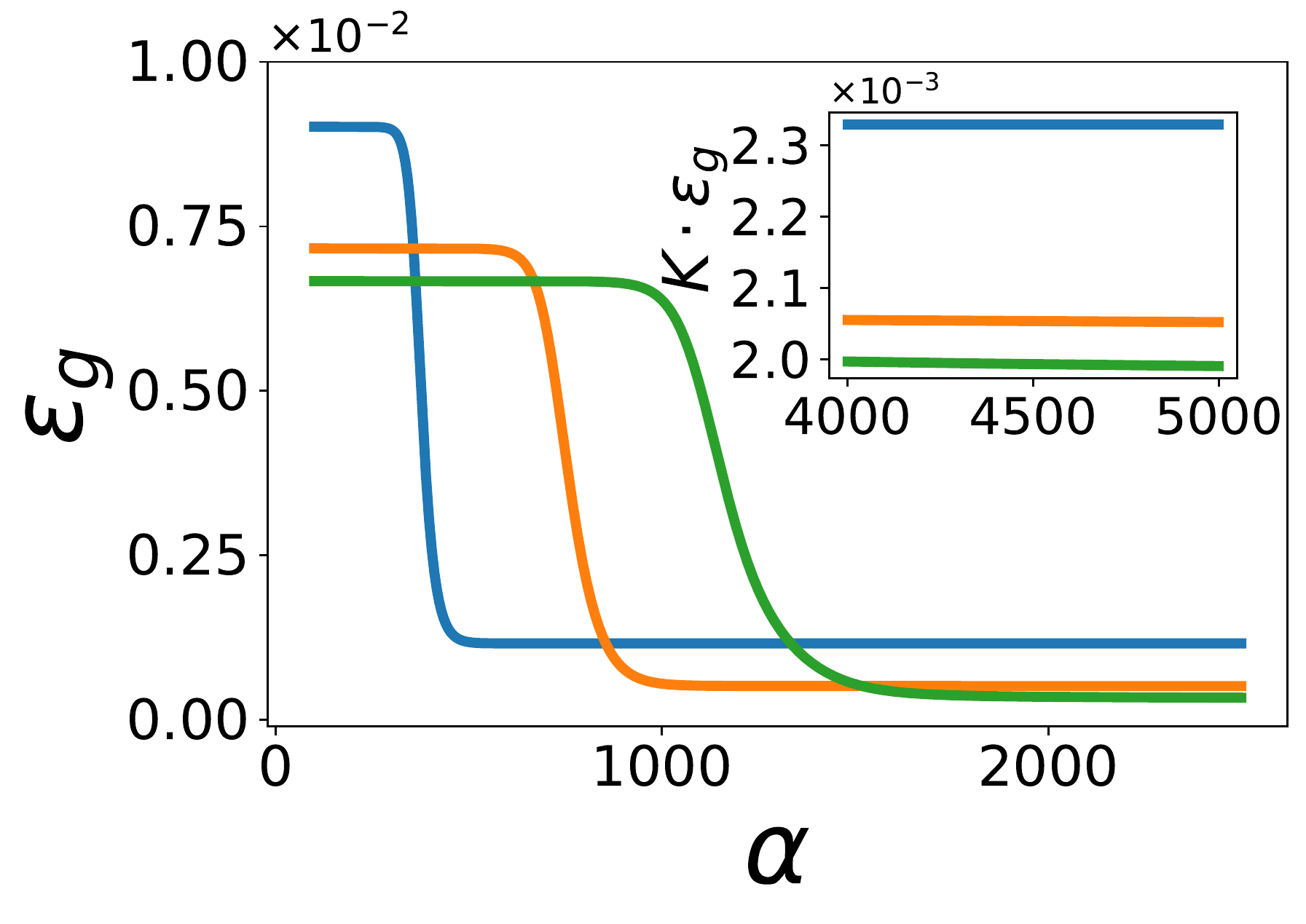}
	\caption{\label{Fig:eps2}   Evolution of the generalization error in the case of  a noisy teacher $ M=2$ and noise variance $\sigma_{\gamma}^2=0.01$ for $K \in\{2,4,6\}$ and $\eta=1$. The asymptotic generalization error decreases with increasing $K$. The inset shows that the  rescaled asymptotic generalization error $K\cdot\epsilon_g$ is only weakly dependent on $K$.}
\end{figure}

So far, we have focused on the learning dynamics  in the over-realizable regime as compared to the realizable one.  However, in the case of a noisy teacher 
it is interesting to also study how the asymptotic generalization error depends on the degree of over-parametrization. Specifically, we consider a noisy teacher 
with output $\zeta_{\gamma}(\boldsymbol{B},\boldsymbol{\xi})=\zeta(\boldsymbol{B},\boldsymbol{\xi})+\gamma$, where random noise $\gamma\in\mathcal{N}(0,\sigma_{\gamma}^2)$ is added to the output of the teacher network. The dependence  of the asymptotic generalization error on $K$ is shown in Fig.~\ref{Fig:eps2}. It is consistent with the scaling $\epsilon_g^{\infty}\propto\frac{1}{K}$ suggested in \cite{goldt2019dynamics} for the case that both layers of the network were fully trained. We thus demonstrate that learning of the second layer is not required for an improved asymptotic generalization error if a suitable normalization is applied to student and teacher outputs. 
 
In conclusion, we have shown that using an appropriate normalization of teacher and student outputs in a two-layer soft committee machine, all student nodes can learn in the over-realizable regime. The generalization error stays finite in the limit of a large number of hidden units, and the value of the symmetric plateau is independent of the number of student nodes. Following the specialization transition, groups of students will imitate one teacher node, with a power-law approach to perfect learning due to the presence of soft modes. In the case of a noisy teacher, the asymptotic generalization error is lower in the 
over-realizable regime as compared to the realizable case.


\end{document}